\documentclass[aps,prb,twocolumn, showpacs,amsmath,amssymb,amsthm]{revtex4-2}
\usepackage{dcolumn}
\usepackage{bm}
\usepackage{graphicx}
\begin{document}
	\title{Three-dimensional ferrimagnetic ground state of triangular-lattice system Ca$_{3}$Co$_{2}$O$_{6}$}
	\author{Santanu De, V. R. Reddy and A. Banerjee}
	\affiliation{UGC-DAE Consortium for Scientific Research, University Campus, Khandwa Road, Indore-452001, India.}
	
	\begin{abstract}
		High temperature one-dimensional (1D) ferromagnetic (FM) chains in Ca$_{3}$Co$_{2}$O$_{6}$ spin system are subjected to a magnetic field and temperature induced  first order phase transition (FOPT) to 3D-ferrimagnetic (FIM) ground state with decrease in temperature (T). Weak-FM interaction of third-nearest-neighbor (nn) interchain removes the frustration effect arising from antiferromagnetic (AFM) interactions of first-nn and second-nn interchains in the underlying triangular-lattice resulting a 3D-FIM ordering of 1D FM chains at low T. However, hindered kinetics of FOPT partially masks this tranformation giving rise to coexistence of non-interacting 1D FM chains with 3D-FIM state at low-T. The existence of all these couplings is further confirmed here by random substitution of S = 5/2 magnetic-impurity into the spin chain of original system. It reveals weakening of FM interactions of both intrachain and third-nn surrounding chains respectively without significant modulation in the AFM coupling of first-nn and second-nn interchains. Thus, influence of AFM interactions is enhanced as compared to effective FM coupling with increase of S = 5/2 impurity content resulting instability of 3D long-range FIM state at low T.
	\end{abstract}
	\maketitle
	\section{Introduction}
    Study of ground state magnetism of triangular-lattice (TL) antiferromagnet is one of the vibrant research areas in condensed matter physics. It is mainly due to various possible interacting ways of underlying spin in this TL that might give rise to many intriguing outcomes like geometrically frustrated spin state and long-range ordering (LRO) \cite{r1,r2,r3,r4,r5,r6}. A great variety of materials belong to this TL spin system \cite{r1,r2,r3,r4,r5,r6} and one of the important compounds of this family is that Ca$_{3}$Co$_{2}$O$_{6}$ (CCO) \cite{r6,r7,r8,r9,r10,r11,r12,r13,r14,r15,r16,r17,r18,r19,r20,r21,r22,r23,r24,r25,r26,r27,r28,r29,r30,r31,r32,r33,r34,r35,r36,r37,r38}. It provides a great opportunity to study the magnetic properties by incorporating many transition metal ions (3d, 4d and 5d) \cite{r6,r7,r8,r9,r10,r11,r12,r13}. It has rhombohedral structure in R\={3}c space group which constructs ABCABC... type of stacking sequence of honeycomb lattice along c-axis \cite{r6}. Moreover, spin chains consisting of alternating trigonal prism (TP) and octahedra (OCT) sites are running along c-axis of this structure. Earlier study clarifies that crystal electric field variation creates dissimilar spin state of Cobalt ion (Co$ ^{3+} $) at two different sites that means S=2 and S=0 spin state at TP and OCT sites  respectively \cite{r14,r15,r16,r17,r18,r19}. In this spin system, spins sitting at TP site are ferromagnetically (FM) coupled and in contrast to that first and second nearest neighbor (nn) surrounding chains are antiferromagnetically (AFM) coupled in this triangular-lattice system \cite{r20,r21,r22,r23,r24}.
    
    A brief overview of literature suggests a rich magnetic phase diagram of this compound in field (H)-temperature (T) space. At the intermediate high T, spins (S=2) are interpreted to be aligned along chain direction due to strong-FM intrachain interaction and resulting one-dimensional (1D) behavior of this system \cite{r21,r25}. Whereas, these 1D FM chains are ordered in three-dimensional (3D) ferrimagnetic (FIM) state at low T and further, this 3D-FIM state transforms to 3D-FM state at higher H \cite{r18,r19,r25,r26,r27,r28,r29,r30,r31,r32,r33}. However, there exists many controversies in this TL Ising system and one of the intensely debated issue is the magnetic ground state. Some earlier studies on CCO compound suggest existence of a FIM like state at low T \cite{r18,r19,r25,r26,r27,r28,r29,r30,r31,r32,r33} whereas few recent reports infer spin density wave like ordering of low T state which slowly transforms to commensurate antiferromagnetic state \cite{r20,r21,r22,r23,r24,r34,r35,r36,r37,r38}.
	
	An attempt is made here to resolve	the nature of three-dimensional ground state by the measurement of dc-magnetization and first \& second order response of magnetic ac-susceptibility at low H in this TL system. It is presented here that this spin system experiences a broad first order phase transition (FOPT) to 3D-FIM state with decreasing T which is supposed to be true ground state. However, hindrance to the transformation kinetics of FOPT causes co-existence of high T state of FM chains and 3D-FIM state at low temperature. The microscopic analogue of this interplay of FOPT and hindered kinetics of FOPT with decrease in T has been described by simultaneously taking into account the strong-FM intrachain coupling and AFM interactions of first-nn \& second-nn along with weak-FM coupling of third-nn interchain in underlying TL. Further, ferromagnetic exchange couplings are tuned with chemical substitution of S = 5/2 (Fe$ ^{3+} $) magnetic-impurity without considerably modifying AFM interactions. Effective FM interaction is reduced in presence of S = 5/2 impurity which leads to domination of AFM couplings and weakens the stability of low T 3D-FIM state.

	\begin{figure*}[tbph]
		\centering
		\includegraphics[width=18cm]{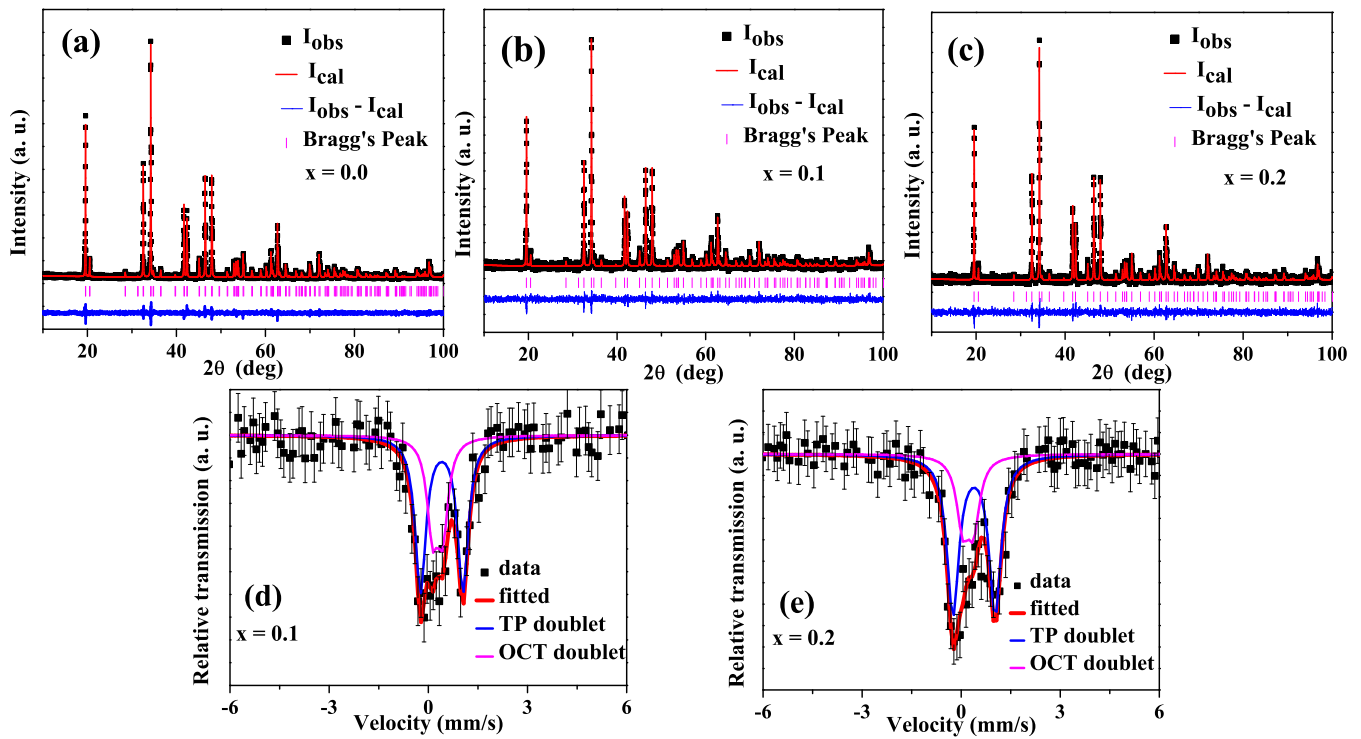}
		\caption{(Color online) (a)-(c) Refined X-ray diffraction pattern of polycrystalline Ca$_{3}$Co$_{2-x}$Fe$_{x}$O$_{6}$ (x = 0.0, 0.1, 0.2) at room temperature and (d)-(e) M\"{o}ssbauer spectrum for x = 0.1 \& 0.2 respectively at room temperature.}
		\label{figs:1}
	\end{figure*}
	
	\section{Experimental}
	We have used same single crystal of CCO as discussed in reference 32. On the other hand, all poly-crystalline compounds, Ca$_{3}$Co$_{2-x}$Fe$_{x}$O$_{6}$ (x = 0.0, 0.1 and 0.2), are synthesized in conventional solid state method. Stoichiometric amount of CaCO$ _{3} $, Co$ _{3} $O$ _{4} $ and Fe$ _{2} $O$ _{3} $ powders for all compositions were separately grinded for 6 hrs and thereafter, the mixtures were heated for 4 days in air at 900$ ^{0} $C in powder form. Finally, they were pelletized after intermediate grinding and heated for another few days followed by slow cooling process down to room temperature. Figures 1 (a) - (c) illustrate room temperature powder X-ray diffraction (PXRD) pattern for all samples. Reitveld refinement of PXRD data show that all compounds are crystallized in rhombohedral structure in R\={3}c space group and within detection limit, there are no extra impurities. The variation of all structural parameters obtained from Reitveld refinement of PXRD data are shown in TABLE I with iron substitution. It is apparent that lattice parameters are not significantly modified after Iron substitution at Cobalt sites. Moreover, it is rather interesting that Iron ions are randomly substituted in both OCT and TP sites for both x = 0.1 and 0.2. Although this random distribution of Iron at Cobalt sites is contradictory to some earlier reports on lower concentration of Iron substitution at Cobalt sites but it becomes similar to that of our result with increasingly higher concentration of Iron \cite{r7,r8,r9}.  
	\begin{table}
		\centering
		\caption{Structural parameters obtained from Reitveld refinement of PXRD data (where (X,Y,Z) is atomic positions in the underlying space group and lattice parameters are found to be a$\sim$9.074 $ \AA $ \& c$\sim$10.337 $ \AA $ in all systems):}
	\begin{tabular}{c|ccccc}
		\hline
		Sample (x)   & atom & X & Y & Z & occupancy\\ 
		\hline \hline 
		&   Ca & 0.369 (6) & 0 & 0.25 & 0.326 (8)\\ 
		0.0  & Co1 & 0 & 0 & 0 & 0.108 (7) \\
		&   Co2 & 0 & 0 & 0.25 & 0.108 (7) \\
		&    O & 0.178 (5) & 0.025 (3) & 0.113 (3) & 0.650 (6) \\
		\hline
		&   Ca & 0.368 (1) & 0 & 0.25 & 0.382 (0)\\ 
		& Co1 & 0 & 0 & 0 & 0.120 (5) \\
		0.1 & Co2 & 0 & 0 & 0.25 & 0.120 (4) \\
		&  Fe1 & 0 & 0 & 0 & 0.007 (5) \\
		&  Fe2 & 0 & 0 & 0.25 & 0.008(8) \\
		&  O & 0.181(5) & 0.026 (1) & 0.113 (5) & 0.809(6) \\
		\hline
		&  Ca & 0.369 (3) & 0 & 0.25 & 0.358 (7)\\ 
		&  Co1 & 0 & 0 & 0 & 0.107 (3) \\
		0.2  & Co2 & 0 & 0 & 0.25 & 0.107 (8) \\
		&   Fe1 & 0 & 0 & 0 & 0.010 (4) \\
		&   Fe2 & 0 & 0 & 0.25 & 0.012 (0) \\
		&  O & 0.180 (0) & 0.025 (6) & 0.113 (5) & 0.728 (3) \\	
		\hline 
	\end{tabular} 
	
\end{table}
	At the same time, distribution of Fe-ions along chain and their charge \& spin states at six coordination are determined from the room temperature M\"{o}ssbauer spectroscopy which are respectively shown in figures 1 (d) and (e). Fitting of M\"{o}ssbauer data with two well defined doublets suggests that random distribution of Fe-ions is observed at both sites which is similar to that of PXRD. Average percentage of Fe-ions at OCT \& TP sites are observed around 40\% \& 60\% for x =0.1 and 30\% \& 70\% for x = 0.2 respectively. The obtained value of isomer shift and quadruple splitting (x = 0.1: TP - 0.41 and 1.28, OCT - 0.28 and 0.31 \& x = 0.2: TP - 0.39 and 1.28, OCT - 0.21 and 0.33) indicate that Fe$ ^{3+} $ ions are in high spins state (S = 5/2) at both sites \cite{r7,r8,r9,r39,r40}. 
	
	Field dependent magnetization measurements parallel to c-axis of single-crystalline CCO were carried out in commercial 16 T Vibrating-Sample-Magnetometer (VSM) whereas the variation of magnetization as a function of temperature and time were recorded in Superconducting-Quantum-Interference-Device (SQUID) magnetometer, MPMS 3, QD, USA. Standard protocols like zero-field-cooled (ZFC), field-cooled-cooling (FCC) and field-cooled-warming (FCW) are used in these measurements and in addition to that many other protocols are designed for various purposes. Similarly, all temperature and field dependence of dc-magnetization in Ca$_{3}$Co$_{2-x}$Fe$_{x}$O$_{6}$ (x = 0.0, 0.1, 0.2) compounds were performed in same 16 T VSM. The linear and non-linear ac-susceptibilities of these poly-crystalline systems were measured in a home-made instrument that can go down to liquid helium temperature from 300 K \cite{r41}. 
	
	\section{Result and Discussions}
	\subsection{Three-dimensional ferrimagnetic ground state and the interplay of first order phase transition \& its hindered kinetics in this triangular-lattice system:}
	
	Temperature dependence of dc-magnetization at 0.5 kOe in zero-field cooled (ZFC), field cooled cooling (FCC) and field cooled warming (FCW) modes along c-axis of single-crystalline Ca$_{3}$Co$_{2}$O$_{6}$ is presented in figure 2. FCC and FCW/ZFC, M (T) curves, demonstrate a thermal hysteresis in the range, T $ \sim $ 12-25 K which is discussed earlier in refs. [31, 32]. Such type of hysteric transformation suggests a broad first-order phase transition. It has already been interpreted that this spin system has a one-dimensional character at high T (25 K$<T<$120 K) resulting from ferromagnetically coupled spins along chain direction \cite{r31,r32}. Therefore, this high T 1D chains transform to 3D-FIM ground state with decreasing T at lower field. This low T state comprises a large uncompensated magnetization whereas recent neutron diffraction and X-ray scattering studies suggest that ground state has no uncompensated spontaneous magnetization \cite{r20,r21,r22,r24,r34,r35,r36,r37,r38}. Moreover, ZFC M (T) curve bifurcates from the path followed in FCC and FCW modes in H-T space. This irreversibility signifies that high T 1D chains partially convert into 3D-FIM state. It ultimately results in co-existence of equilibrium 3D-FIM phase with 1D FM chains at low T \cite{r31,r32}. There are many systems including half doped manganite, heusler and shape memory alloys etc. which undergo a similar type of broad FOPT \cite{r42,r43,r44,r45,r46,r47,r48}. The arrest of transformation kinetics associated with FOPT in these systems leads to co-existence of thermodynamically equilibrium and glass like non-equilibrium magnetic states at low temperature. Thus, the interplay of FOPT and hindered kinetics of FOPT may lead to such an interesting phase co-existence scenario in this spin chain system. Furthermore, unlike other systems, the growth process of low T 3D-FIM ordering in this spin system has to be described by considering various possible interactions in the underlying triangular-lattice.
	
	\begin{figure}[htbp]
		\centering
		\includegraphics[width=8.5cm]{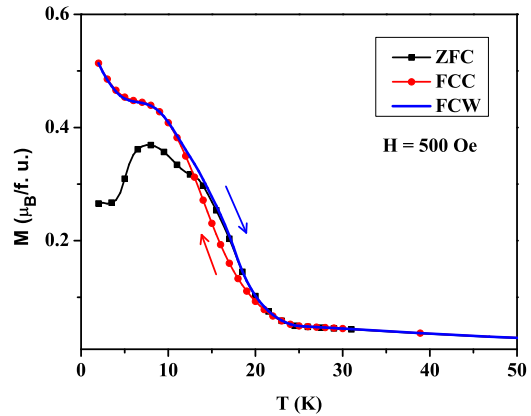}
		\caption{(Color online) Variation of magnetization as a function of temperature along the c-axis of CCO single-crystal. This measurement was performed in the presence of 0.5 kOe field in three different protocols (ZFC, FCC and FCW).}
		\label{fig:1}
	\end{figure} 
	
	\begin{figure*}[htbp]
		\centering
		\includegraphics[width=18cm]{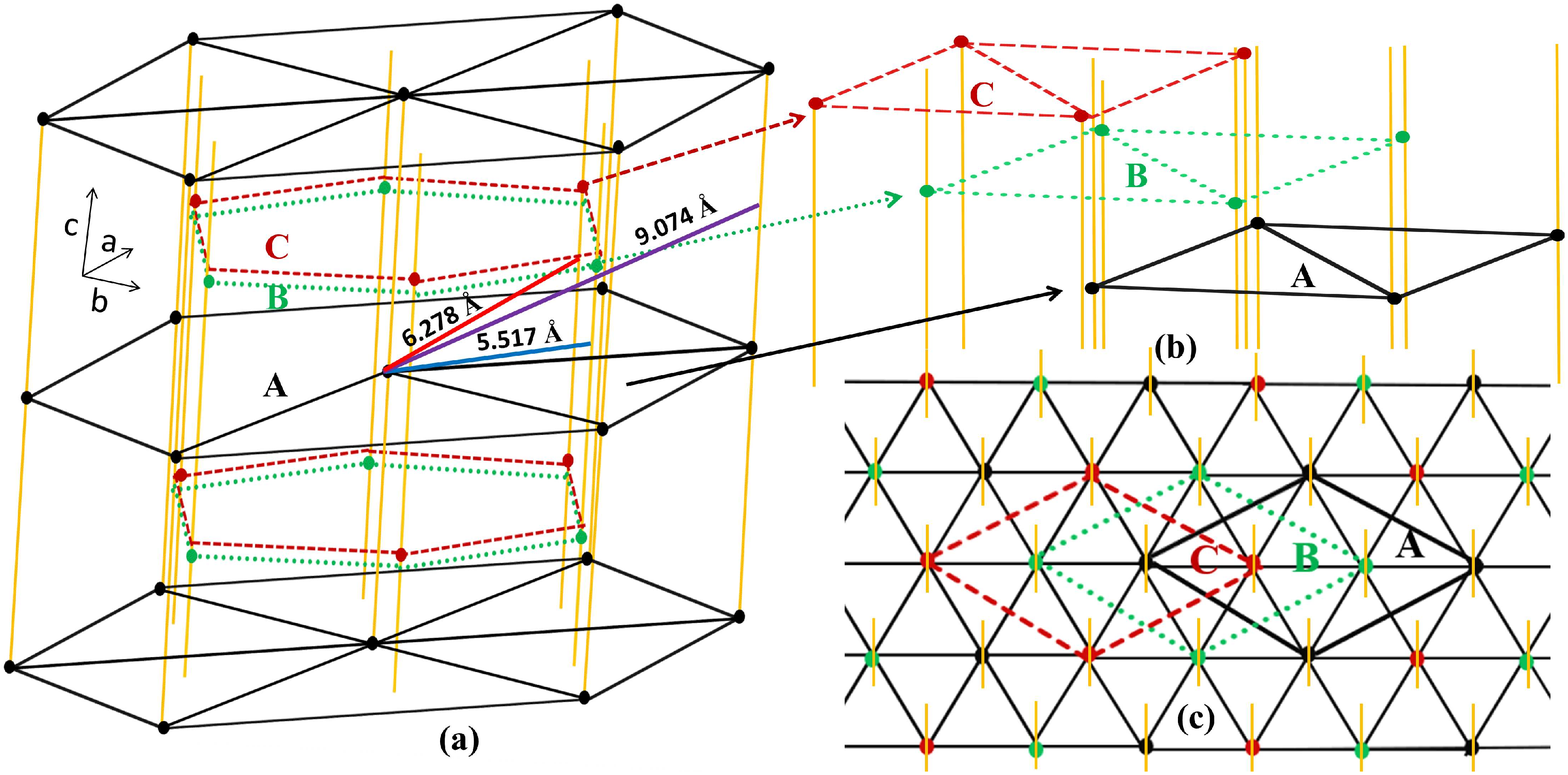}
		\caption{(Color online) (a) The schematic diagram of ABC-stacked honeycomb lattice where yellow solid lines correspond to linear spin chain and black solid, green dotted and red dashed lines stand for A, B and C layers respectively of hexagonal lattice. (b) The stacking arrangement of a/a rhombic cell along crystallographic c-axis in A (black solid line), B (green dotted line) and C (red dashed line) layers respectively of ABC-stacked honeycomb lattice. (c)  Projection of this crystal structure in ab-plane.}
		\label{figure:2}
	\end{figure*}
	
	The growth mechanism of 3D-FIM state at low T is successfully interpreted by taking into account interactions of first-nn, second-nn and third-nn surrounding FM chains in this TL \cite{r1}. Figure 3 (a) shows ABC-stacked honeycomb lattice schematically where yellow solid lines are the spin chains along crystallographic c-axis and A, B and C layers are respectively illustrated by black solid lines, green dotted lines and red dashed lines. The blue, red and violet solid lines are the first-nn (5.517 $\AA$), second-nn (6.287 $\AA$) and third-nn (9.074 $\AA$) surrounding chains. To simplify this structural complexity, a sub-lattice of dimension $ \sim $ a/a is considered here that preserves hexagonal symmetry of original crystal structure ($ \textquoteleft $a' corresponds to lattice parameter in ab-plane which is $\sim$ 9.074 $\AA$). Figure 3 (b) illustrates the stacking sequence of this a/a sub-lattice along crystallographic c-axis by black solid lines, green dotted lines and red dashed lines in A, B and C layers respectively of original structure. This is further simplified in figure 3 (c) by projecting this structure in ab-plane. Now, three successive sub-lattices (A, B and C) of dimension a/a as shown in figure 3 (c) are considered to interprete the ferrimagnetic ordering in this spin system. It is evident that each sub-lattice consists of four similar spin chains which passes through the vertices of trianglar-unit. The ferrimagnetic ordering can be established in these sub-lattices if eight spin chains in two of these sub-lattices are aligned in the same direction and remaining one sub-lattice consisting four spin chains becomes opposite to that of spin chains ordering in other two sub-lattices. In this way, three different arrangements of spin chains give rise to same ferrimagnetic state. So, it indicates that this ferrimagnetic ground state is a three folds degenerate state.
	
	One of the ferrimagnetically ordered states is schematically depicted in figure 4 where it is assumed that (1) one-dimensional ferromagnetic chain plays role of a giant moment (arrow symbol) due to strong nature of FM intrachain coupling compared to average interchain interaction and (2) AFM interaction strength of first-nn (J$ ^{1} _{\perp} $) is comparable to the second-nn (J$ ^{2} _{\perp} $) due to their close bond lengths in this system and (3) weak-FM coupling of all third-nn interchains \cite{r21,r22,r23}. It is apparent that small perturbation caused by the third-nn ferromagnetic interaction (J$ ^{3}_{\perp} $) along with AFM J$ ^{1} _{\perp} $ and J$ ^{2} _{\perp} $ couplings stabilize this 3D-FIM ordering in this predominantly antiferromagnetically coupled triangular-lattice system. Because weak-FM coupling of third-nn interchain would remove the tendency of TL frustration arising from first-nn and second-nn AFM couplings of giant moments or 1D FM chains. It results in the ferromagnetic alignment of five AFM interchain bonds in a ABC-stacked rhombic cell of dimension a/a as depicted by purple dotted ellipses in figure 4. Therefore, it can be considered that the energy cost for these ferromagnetic alignments will be given by all weak-FM third-nn interchain bonds in a ABC-stacked lattice. The energy required for the ferromagnetic ordering of one of the AFM interchain bonds is 2J$ ^{1} _{\perp} $. Thus, total energy cost of ferromagnetic ordering of five AFM interchain bonds (see the purple dotted ellipses in figure 4) is 10J$ ^{1} _{\perp} $. At the same time, there are 12 weak-FM third-nn interchain bonds that ferromagnetically couples the underlying four giant moments in each individual layer. These bonds are shown in figure 4 by green dotted lines in one of layers (B layer) of ABC-stacked lattice. So, the total energy of the weak-FM interaction of all third-nn 1D FM chains/giant moments in a ABC stacked layer is 36J$ ^{3} _{\perp} $. So, it is found that 10J$ ^{1} _{\perp} $ $ \sim $ 36J$ ^{3} _{\perp} $ and hence, the ratio of $ \dfrac{J^{3} _{\perp} }{J ^{1} _{\perp} } $ becomes
	\begin{equation*}
	\dfrac{J^{3} _{\perp} }{J ^{1} _{\perp} } = 0.28 ................(1)
	\end{equation*}
	Moreover, previous calculation of average interchain coupling strength using onset temperature of LRO in mean field approximation \cite{r25} suggests that 
	\begin{equation*}
		({J ^{1} _{\perp} } + {J ^{2} _{\perp} } + {J ^{3} _{\perp} }) = 1.04 .........(2)
	\end{equation*}
	 For J$ ^{1} _{\perp} $ $\sim $  J$ ^{2} _{\perp} $, equation (2) becomes 
	 \begin{equation*}
	 	(2{J ^{1} _{\perp} } + {J ^{3} _{\perp} }) = 1.04 ...........(3)
	 \end{equation*}
     Therefore, the individual values of J$ ^{1}_{\perp}$, J$ ^{2}_{\perp}$ and J$ ^{3}_{\perp} $ can be estimated from eqs. (1) and (3) which are observed to be around J$ ^{1}_{\perp}$ $\sim$ J$ ^{2}_{\perp}$ = - 0.45 K and J$ ^{3}_{\perp} $ = 0.14 K respectively. Although, significant contribution of orbital moment to total magnetization is observed in this spin system but all calculations are performed here by assuming only spin moment without the loss of generality. The resultant magnetization of this ferrimagnetically ordered state is found to be 1/3 times of saturation moment of fully magnetized ferromagnetic state where out of 12 spins, eight in A and C sub-lattices are oriented in up and remaining four spins in B sub-lattice are aligned in down direction in a ABC-stacked triangular-lattice. Therefore, 1D FM chains undergo a phase transition to a three-dimensional ferrimagnetic state with decrease in temperature which is of first order. It occurs due to influence of both intrachain and interchain couplings at low temperature as proposed in ref. [25]. However, hindered kinetics of FOPT partially masks this transformation that leads to co-existence of 1D FM chains with 3D-FIM state at low T. 
	\begin{figure}[htbp]
		\centering
		\includegraphics[width=8.5cm]{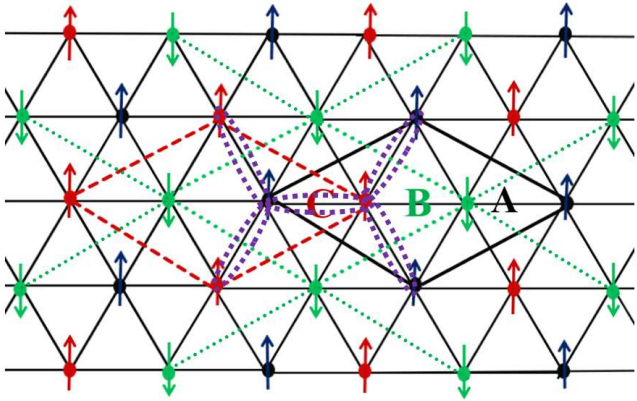}
		\caption{(Color online) Ferrimagnetic ordering in the projected ab-plane of ABC-stacked honeycomb lattice.}
		\label{fig:2}
	\end{figure}   
	 
	Figure 5 shows both the virgin and envelope curves of M (H) which are measured along c-axis of CCO single-crystal at 10 K in the range $ \pm $ 20 kOe. It is found that virgin curve in the field increasing (FI) cycle lies outside of the envelope with lower magnetization value and there is a small opening in the field decreasing (FD) \& subsequent FI branches of envelope curve. Existence of randomly oriented high T 1D FM spin chains is responsible for the lower value of magnetization in virgin path which converts into 3D-FIM state with field excursion. The whole system is ordered in 3D-FIM state above 15 kOe which is confirmed by observed magnetic moment $ \sim $1.7 $ \mu_{B} $/f.u. ($ \sim $1/3M$ _{s} $, where M$ _{s} $ is the saturation moment which is found to be at about $ \sim $ 5 $ \mu_{B} $/f.u. for parallel orientation of the single-crystal) \cite{r32}. On the other hand, it remains almost in 3D-FIM state in FD branch of envelope whereas partial conversation to the 1D FM chains occurs in next FI cycle, as a result, FD and FI branches of envelope do not coincide. Hence, M (H) at 10 K implies the presence of high T 1D FM chains with equilibrium 3D-FIM phase due to interrupted nucleation and growth process of FOPT that hinders a fraction of the 1D FM coupled spin chains to take part in the transformation to 3D-FIM state in underlying TL. Similar observation is also found in poly-crystalline CCO, however, phase fractions are different with respect to single-crystal due to random orientation of magneto-crystalline anisotropy of individual grain in poly-crystal \cite{r31}. Experimental data have also been analyzed in context of a classical model to support qualitative understanding of this phase co-existence. An expression was used earlier to describe two plateaux in M (H) curve of single-crystalline CCO by Hardy et al. \cite{r49} which is given below
	\begin{equation*}
		M_{\parallel}(H) = M_{s}[\dfrac{1}{3}(1-e^{-\frac{H}{a}})
		+\frac{2}{3}(\dfrac{1}{1+e^{-\frac{(H-H_{c_{1})}}{b}}})].....(4)
	\end{equation*}
	Where M$ _{s} $, a \& b and H$ _{c_{1}} $ correspond to saturation magnetization, adjustable parameters and critical field for 3D-FIM to 3D-FM transition respectively. First term in equation (4) represents magnetization process of 3D-FIM state at low T whereas second one corresponds to 3D-FM state which dominates above H$ _{c_{1}} $. Thus, in low field range equation (4) can be written as
	\begin{equation*}
		M_{\parallel}(H) = \dfrac{M_{s}}{3}(1-e^{-\frac{H}{a}}) for H<H_{c_{1}}  ......(5)
	\end{equation*}
	\begin{figure}[htbp]
		\centering
		\includegraphics[width=8.5cm]{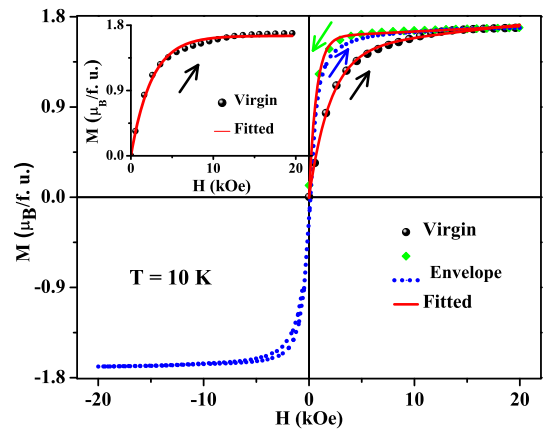}
		\caption{(Color online) Virgin and envelope curves  of field-dependent magnetization at 10 K in the field range $ \pm $ 20 kOe whereas the inset shows only virgin curve of M (H). All solid lines in this figure represent fitting of experimental data with a classical model.}
		\label{fig:2}
	\end{figure}
	An effort is made to fit the virgin curve of M (H) with this equation which is shown in the inset of figure 5 and a slight deviation is observed here due to exclusion of the transformation process of 1D FM chains to 3D-FIM state. However, it has been modified by taking into account this conversion process along with the magnetization process of 3D-FIM state and represented in following eq. (6) where M (H) behavior of 1D FM chains is considered to be linear in low H limit as that of non-interacting paramagnet. 
	\begin{equation*}
		M_{\parallel}(H) = M_{s}[\dfrac{1}{3}(1-e^{-\frac{H}{a}})+cH] for H<H_{c_{1}}  ......(6)
	\end{equation*}
	In equation (6), c stands for the fraction of high T 1D FM chains. This equation (6) is used to fit the FI cycle of virgin curve and the subsequent FD cycles as shown in figure 5. Obtained value of coefficient c from the fitting of M-H with eq. (6) is found 92\% lower in FD branch of envelope with respect to FI cycle of virgin curve at 10 K which confirms that the system prefers to stay in 3D-FIM state in FD cycle of envelope curve.
	
	Therefore, it is obvious from both M (T) and M (H) that high T 1D FM chain coexists with equilibrium 3D-FIM phase over a small range of temperature and field in this spin system due to the interplay of thermodynamic FOPT and its hindered kinetics. This interplay is tuned here by cooling this spin system in various magnetic fields at a fixed T which is expected to initiate the growth of low T 3D-ferrimagnetic state. Figure 6 (a) demonstrates the variation of magnetization with field at 10 K after cooled in presence of different cooling fields (H$ _{cf} $). 
	\begin{figure}[htbp]
		\centering
		\includegraphics[width=8.5cm]{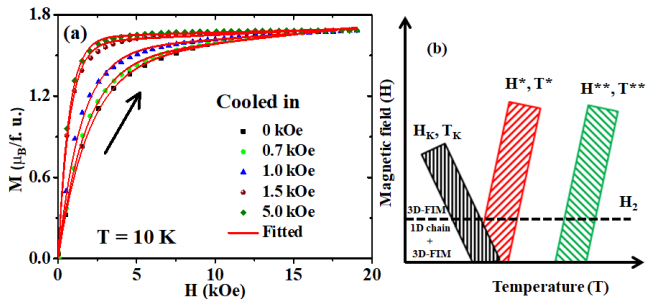}
		\caption{(Color online) (a) M (H) at 10 K which are measured only in the first quadrant. Before each measurement, this spin system was cooled at various cooling fields from 200 K and then magnetic field was linearly switched off. Solid line demonstrates the fitted curve of this magnetization process in context of a classical model (b) Schematic of an intuitive H-T phase diagram}
		\label{fig:2}
	\end{figure}
	In this protocol, the system was cooled down to 10 K at different H$ _{cf} $ and before data recording in field increasing branch the H$ _{cf} $ was isothermally reduced to zero. It is observed that magnetization value enhances with increasingly higher H$ _{cf} $ until this system reaches to 3D-FIM state. Finally, it suddenly moves to one third times of total saturation moment ($ \sim $1.7 $ \mu_{B} $/f.u.) above a certain cooling field which indicates complete FIM ordering in three-dimension. Thus, field cooling process initiates the first order transition of high T 1D FM chains to 3D-FIM state which results in decrease of 1D chain fraction at low T. Obtained initial phase fractions of 1D FM chains are approximately found to be around 80.3\%, 69.2\%, 45.7\%, 15.1\% after cooled in 0, 0.7, 1.0, 1.5 kOe fields respectively. These phase fractions were calculated from the observed difference in magnetization at 0.5 and 18 kOe fields with respect to the magnetization of 3D-FIM state (1.7 $ \mu_{B} $/f.u.) in all cases. It is further supported by estimated value of coefficient, c, of eq. (6) at various H$ _{cf} $ which is obtained from fitting of these experimental data with eq. (6). It decreases for higher cooling field and finally, it becomes negligibly small at about H$ _{cf} $$ \sim $ 5 kOe. Such a remarkable outcome is interpreted in context of H-T phase diagram for a broad and hindered first-order phase transformation process \cite{r45,r46,r47,r48}. A schematic representation of phase co-existence in H-T space is shown in figure 6 (b) where (H**, T**), (H*, T*) and (H$ _{K} $, T$ _{K} $) corresponds to superheating, supercooling and kinetic arrest bands respectively and H$ _{2} $ is the characteristic field. There is co-existence of both 1D FM chains and 3D-FIM state at low T when H$ _{cf} $$ < $ H$ _{2} $ due to partial overlapping of both supercooling and kinetic arrest bands. An increase of H$ _{cf} $ toward H$ _{2} $ lessens the fraction of high T 1D chains because larger area of supercooling band is available to the system in this path which initiates undergoing FOPT process. On the other hand, cooling in a field which is above H$ _{2} $ leads to complete transformation of high T 1D chains to stable 3D-FIM phase because these paths initially meet supercooling band compared to previous case and hence almost no high T phase is found while cooled this system in such path. The magnetization value of 3D-FIM state after complete removal of high T 1D FM chain fraction is consistent with theoretical value. Thus, this new type of phase co-existence strongly depends on the path followed in H-T space.
	
	Time evolution of magnetization is further investigated to understand the nature of this high T phase across hindered FOPT. M (t) along c-axis of single-crystal at various temperatures are measured in presence of 0.5 kOe field. Before each measurement the system was cooled down to respective measuring temperature at same 0.5 kOe field from 200 K.      
	\begin{figure}[htbp]
		\centering
		\includegraphics[width=8.5cm]{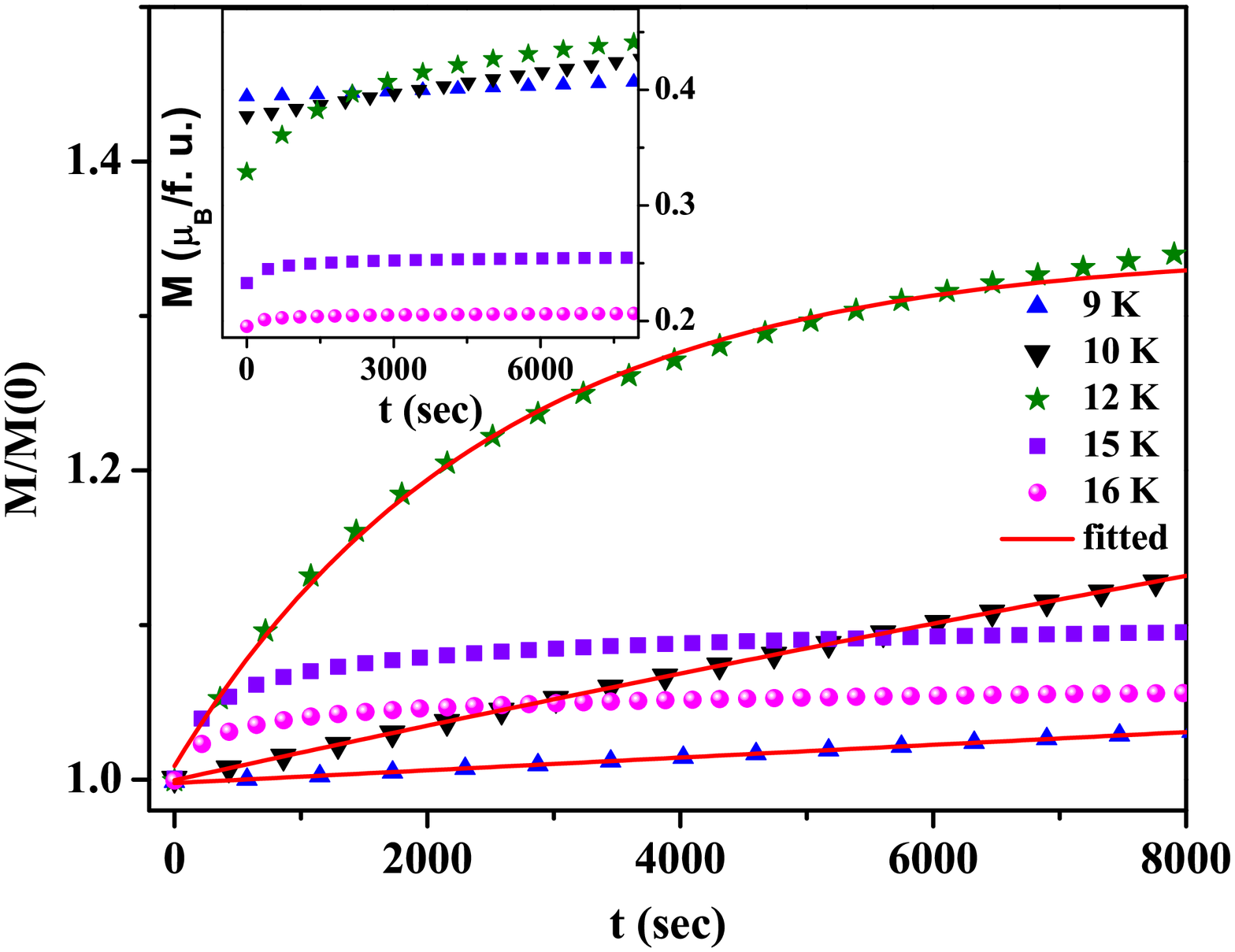}
		\caption{(Color online) Time dependence of magnetization at different temperatures in presence of 0.5 kOe field along crystallographic c-axis. After normalizing the vertical scale with initial value of magnetization, it is fitted with a stretch exponential function as shown by solid lines. Whereas, the inset shows time dependence of magnetization in absolute scale. Before each measurement the system was cooled in 0.5 kOe field down to the respective measuring temperature.}
		\label{fig:2}
	\end{figure}
	Figure 7 and the inset illustrate normalized M (t) data with respect to initial magnetization and M (t) in absolute scale respectively at different temperatures.   
	\begin{figure*}[htbp]
		\centering
		\includegraphics[width=18cm]{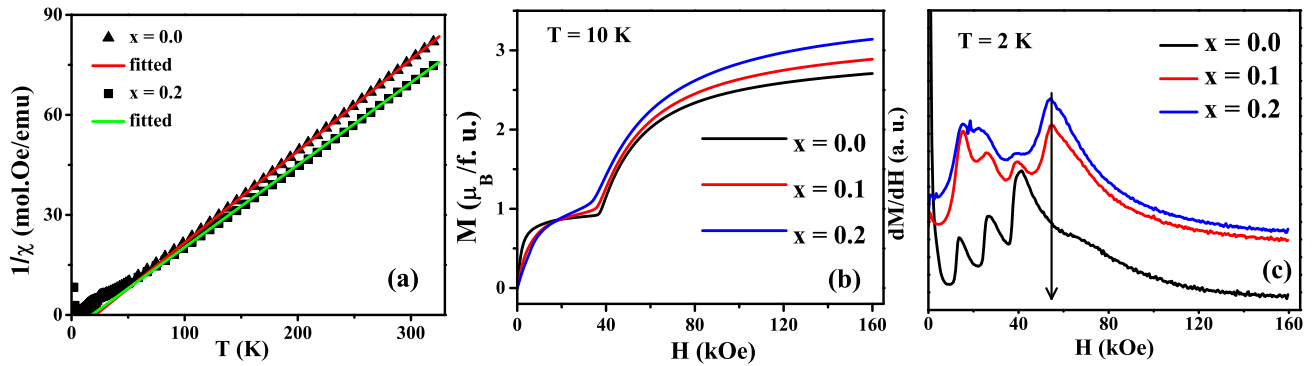}
		\caption{(Color online) (a) Variation of inverse dc-susceptibility (H/M) as a function of temperature where solid line represents Curie Weiss fitting at high T. Here, x = 0.1 is not shown for the shake of clarity., (b) Field dependent magnetization at 10 K, (c) The plot of first order derivative of magnetization with respect to applied field vs magnetic field in all poly-crystalline systems, Ca$_{3}$Co$_{2-x}$Fe$_{x}$O$_{6}$ (x = 0.0, 0.1 and 0.2).}
		\label{fig:2}
	\end{figure*}
	It is evident that the magnetization rapidly increases as a function of time with increase in measuring temperature from 9 to 12 K. It indicates faster kinetics of 1D FM chains at high temperature which is similar to that of glass like behavior. It might arise due to the negligible hight of the free energy barrier associated with FOPT process in this temperature range \cite{r43,r44,r48}. Faster nature of kinetics is also supported by estimated value of relaxation time ($ \tau $) which varies from 10$ ^{7} $ to 10$ ^{3} $ in this temperature range. It is estimated from the fitting of experimental M (t) data with stretched exponential function, M (t) = M (0) + $ \delta $M exp(-t/$ \tau $), where M (0) and $ \delta $M are being the initial magnetization value and magnetization change in the measuring time scale. Whereas, the magnetic relaxation process again slows down at further higher measuring temperature (near about superheating limit). Because height of the free energy barrier enhances at higher measuring temperature which hinders 1D FM to 3D-FIM transformation process \cite{r43,r44,r48}. Hence, a glass like dynamics of high T 1D FM chains is observed while it persists below the supercooling limit of FOPT \cite{r32}.

	\subsection{Magnetic dilution of 3D-ferrimagnetic ground state with S = 5/2 magnetic impurity (Fe$^{3+}$) substitution in this triangular-lattice spin system:}
	Presence of AFM exchange interactions of first-nn \& second-nn and weak-FM coupling of third-nn surrounding FM chains in the 3D-FIM state at low T are shown from the chemical substitution of S = 5/2 magnetic-impurity in original CCO system. This is supposed to be a conventional way to modify complex exchange pathways that effectively results in alternation of the magnetic ground state. Figure 8 (a) shows the plot of Curie Weiss fitting of temperature dependent inverse magnetic susceptibility (H/M) at high T range for Ca$_{3}$Co$_{2-x}$Fe$_{x}$O$_{6}$ (x = 0.0 and 0.2) compounds. The effective magnetic moment is found to be increased from 5.1 to 5.6 $ \mu_{B} $ with increase in Iron content from x = 0.0 to 0.2. It indicates the high spin state of Fe$ ^{3+} $ (S = 5/2) at both OCT and TP sites respectively which is shown in the M\"{o}ssbauer study. Moreover, the obtained value of Curie Weiss temperature in these poly-crystals is found to be decreased from 33 - 22 K with increasingly higher concentration of S = 5/2 impurity. Although the effective magnetic moment increases with Iron substitution due to higher spin state of Iron but the decrease of extrapolated Curie Weiss temperature in iso-structural Iron substituted compounds implies that the effective ferromagnetic coupling along crystallographic c-axis is reduced. It might occur due to either non-magnetic like behavior of Iron in 1D FM chain of Cobalt or influence of a  AFM/FM intrachain exchange coupling of Fe$ ^{3+} $ (3d$ ^{5} $) to the Co$ ^{3+} $ (3d$ ^{6} $) via oxygen ligand. It can be understood from the nature of field dependent dc-magnetization for all three compounds at 10 K as shown in figure 8 (b). It shows lower magnetization in the low field range whereas an opposite nature of magnetization is observed above 20 kOe field with increasingly higher content of Iron at Cobalt sites. Such a contrasting behaviour might imply that Fe$ ^{3+} $ (S = 5/2) couples antiferromagnetically to that of neighbouring Co$ ^{3+} $ (S = 2) along the chain in low field range. This ultimately reduces the effective strength of ferromagnetic coupling along the spin chain which is consistent with the decrese of Curie Weiss temperature with Iron substitution. On the other hand, the antiparallely align S = 5/2 spins of Iron along the chain become ferromagnetic with further field excursion resulting an opposite nature of magnetization in the higher field range. Furthermore, the maximum magnetization observed at highest applied field in saturation pathway of these poly-crystals is increased with increase in Iron contents. This is again consistent with the analogy of high spin state of Fe$ ^{3+} $ (S =5/2) in these compounds. The AFM nature of interaction in between Fe$ ^{3+} $ (S = 5/2) and Co$ ^{3+} $ (S = 2) along the chain is further substantiated from the plot of first order field derivative of magnetization vs magnetic field at 2 K which is shown in figure 8 (c) for all poly-crystalline systems. It is obvious that a new magnetization plateau appears at about 55 kOe (see downward arrow of figure 8 (c)) in Iron doped compounds as compared to CCO system. This characteristic field is found to be different with respect to the regularly spaced characteristic fields (15, 27 and 39 kOe) in the lower field range. It might arise due to the ferromagnetic ordering of antiparallelly aligned S = 5/2 spins of Iron along the chain. Therefore, it is confirmed that Fe$ ^{3+} $ (S = 5/2) is antiferromagnetically coupled to the Co$ ^{3+} $ (S = 2) along the chain in S =5/2 impurity substituted systems. So, high T effective spatial correlation changes from 1D FM to 1D FIM with iron substitution. Similar study was done earlier by many groups but it was interpreted that there is a deviation from ideal 1D character at high T due to non-magnetic like behavior of Fe$ ^{3+} $ (3d$ ^{5} $) in similar iron substituted compounds \cite{r9}. Random substitution of Iron at both TP and OCT sites as observed in our case might be responsible for such a contradictory result at lower concentration of Iron. Another intriguing observation is that the characteristic fields (15, 27 and 39 kOe) associated with the regularly spaced magnetization plateaus appearing in lower field range of Iron doped systems remain same as that of CCO spin system. Although, there is a prolong controversy with field induced magnetization plateaux of CCO spin system in previous literature \cite{r6}. However, the AFM coupling of first-nn and second-nn surrounding spin chains have to be considered in order to describe this anomalous behavior because the critical fields are related with these AFM interchain couplings. Thus, it can be considered that these AFM interchain interactions are not affected significantly with incorporation of S = 5/2 magnetic-impurity.  
	
	Figure 9 (a) represents temperature dependence of first order magnetic ac-susceptibility ($\arrowvert \chi_{1}\arrowvert$) at ac field of amplitude 3 Oe and frequency 231.1 Hz in all poly-crystals (x = 0.0, 0.1 and 0.2). In the intermediate high T range, first order ac-susceptibility monotonically increases with decrease in T down to the characteristic temperature (T$ _{LRO} $) in all systems. It might arise due to 1D-FM or effective 1D-FIM correlation in CCO or Iron substituted CCO compounds. Below T$ _{LRO} $, $\arrowvert \chi_{1}\arrowvert$ suddenly rises in these spin systems which is known as onset temperature of long-range 3D-FIM ordering. However, the T$ _{LRO} $ reduces from 25 to 21 K along with decreasing the sharpness of tranisiton with increase in the concentration of S = 5/2 impurity. Decrease of T$ _{LRO} $ with Iron susbtitution indicates the reduction of average interchain coupling. There are three dominating indirect exchange interactions among surrounding chains in this TL system as suggested by proposed intuitive model. It has already been shown from figure 8 (c) that the AFM couplings of first-nn and second-nn surrounding spin chains remain almost unaffected even after Iron substitution in CCO system. Thus, it can be argued that the average interchain coupling strength lessens only due to weakening of the effective interaction strength of third-nn surrounding chains in S = 5/2 impurity substituted system (x = 0.1 and 0.2). So, the long-distant short range exchange interactions of neighbouring spin chains are affected considerably with Iron substitution which is similar to that of many earlier observations \cite{r50}. It could be due to the higher probablity of finding of Fe$ ^{3+} $ (S = 5/2) at the long-distant neighbouring spin chain of Co$ ^{3+} $ (S = 2) as compared to neighbouring S = 2 spin chain for lower concentration of Iron. Therefore, influence of AFM interaction enhances with random substitution of Iron at Cobalt sites due to reduction of the effective strength of both strong-FM intrachain interaction and weak-FM coupling of third-nn interchains. These would ultimately lessen sharpness of the transition occuring below T$ _{LRO} $. So, the instability of 3D-FIM ground state occurs due to decrease in the effective ferromagnetic coupling strength with S = 5/2 impurity substitution. In contrast, a complex ground state scenario in these systems has been described in previous literature \cite{r9,r20,r21,r22,r23,r24,r34,r35,r36,r37,r38}.
	\begin{figure}[htbp]
		\centering
		\includegraphics[width=9cm]{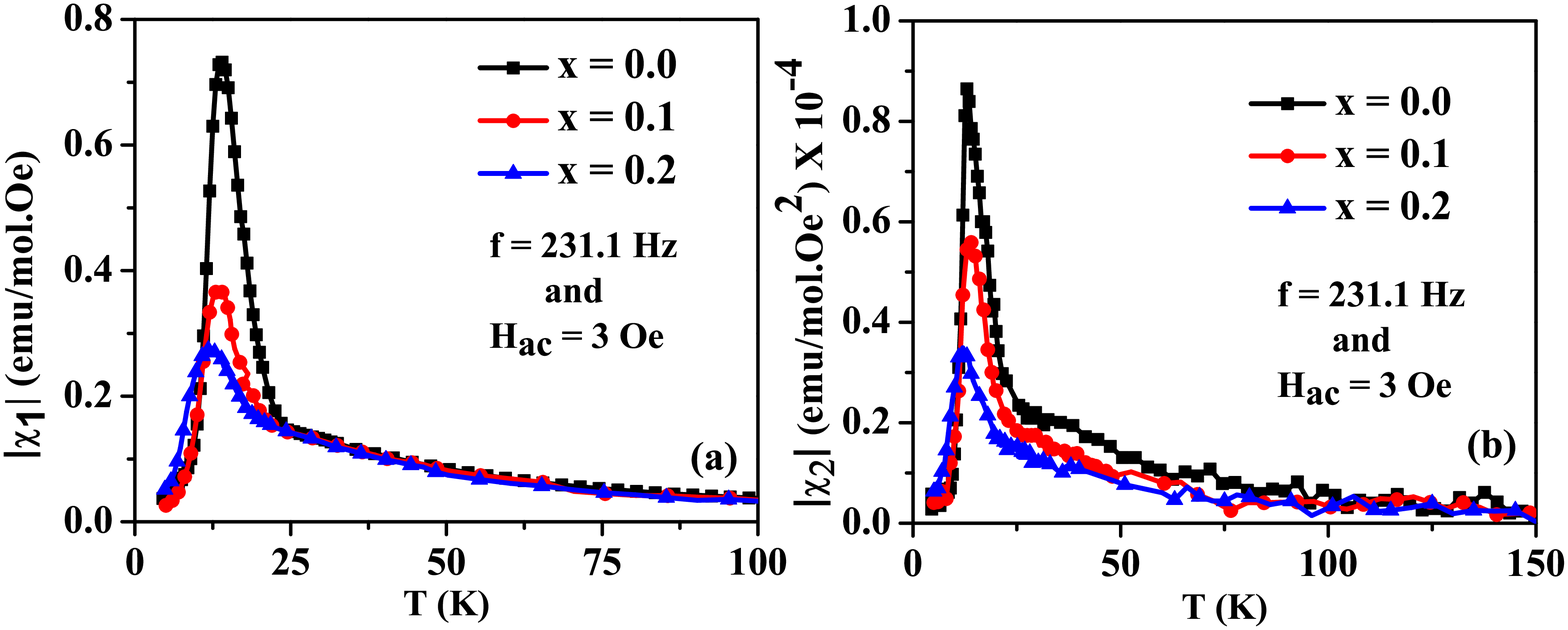}
		\caption{(Color online) Variation of (a) first and (b) second harmonic response of magnetic ac-susceptibility with temperature. All the measurements were performed at 3 Oe ac field of frequency 231 Hz without any dc biasing field.}
		\label{fig:2}
	\end{figure}
	Now, a conclusive evidence is shown here to support the magnetic dilution effect in this ferrimagnet with S = 5/2 impurity substitution which is found in the measurement of second order magnetic ac-susceptibility as a function of temperature at low field. To the best our knowledge, this is the first report of second order response study of magnetic ac-susceptibility in these TL spin systems. This is one of the powerful tools to substantiate true ground state of these systems among many other states of same energy as that of differentiation of spin glass and superparamagnet \cite{r51}. $\arrowvert \chi_{2}\arrowvert$ (T) is illustrated in figure 9 (b) for all poly-crystalline compounds (x = 0.0, 0.1 and 0.2). Similar to $\arrowvert \chi_{1}\arrowvert$, it monotonically increases with lowering T followed by another rapid increase below onset temperature of long-range ordering. Second order susceptibility below 120 K in absence of superimposed dc field suggests presence of an internal symmetry breaking field in these spin systems. It arises due to the ferromagnetic ordering along spin chain at intermediate high T for x = 0.0. Second order susceptibility values are progressively lower in this temperature range for Iron substituted samples due to 1D-ferrimagnetic correlation along the chain \cite{r52}. With decrease in T, these high T states transform to a spontaneously magnetized 3D-FIM state at low T which is of first order. However, suppression of $\arrowvert \chi_{2}\arrowvert$ below T$ _{LRO} $ with increasingly higher concentration of S = 5/2 magnetic-impurity indicates that effective internal field is reduced because of the instability of 3D long-range FIM state. It is due to decrease in the effective FM couplings strength of both intrachain \& third-nn interchain. Therefore, the interaction of strong-FM intrachain and AFM coupling of first-nn \& second-nn interchains along with weak-FM interaction of third-nn interchain appear to stabilize the 3D-FIM ordering at low temperature in CCO spin chain system. 
	
	\section{Conclusions}
	Triangular-lattice system, Ca$_{3}$Co$_{2}$O$_{6} $, undergoes a broad first order 1D-FM to 3D-FIM state transition with decrease of temperature. The growth of this low T 3D-FIM state occurs due to influence of both intrachain and interchain couplings in underlying triangular-lattice. However, this interacting way has been interrupted partially due to hindrance kinetics of FOPT which is responsible for the co-existence of some non-interacting 1D chain with equilibrium 3D-FIM phase at low temperature. Further, instability of 3D-FIM state is shown by chemical substitution of S = 5/2 impurity in original system that arises due to the reduction of FM interactions of intrachain \& third-nn interchain respectively and increase in the domination of first-nn \& second-nn AFM couplings. Thus, it is proved that both intrachain and interchain coupling have an essential role in the growth process of low T FIM state in this Ising spin system.
	
	\section{Acknowledgments}
	We thank Dr. N. P. Lalla for X-ray diffraction and related discussions. Dr. R. Raghunathan and Dr. S. B. Roy are thanked for various iscussions. SD thanks Mr. Biswajit Dutta and Mr. Kranti kumar for their continuous help and support during ac-susceptibility measurement.


\begin{thebibliography}{}
		\bibitem{r1} M. F. Collins and O. A. Petrenko, Can. J. Phys.  \textbf{75}, 605 (1997).
		\bibitem{r2} M. Mekata, J. Phys. Soc. Jpn. \textbf{42}, 76 (1997).
		\bibitem{r3} F. Matsubara and S. Inawashiro, J. Phys. Soc. Jpn. \textbf{56}, 2666 (1987).
		\bibitem{r4} M. Mekata and K. Adachi, J. Phys. Soc. Jpn. \textbf{44}, 806 (1977).
		\bibitem{r5} T. Takagi and M. Mekata, J. Phys. Soc. Jpn. \textbf{64}, 4609 (1995).
		\bibitem{r6} Y. B. Kudasov, A. S. Korshunov, V. N. Pavlov and D. A. Maslov, Phys. Usp. \textbf{55} (12), 1169 (2012).
		\bibitem{r7} H. Kageyama, S. Kawasaki, K. Mibu, M. Takano, K. Yoshimura, and K. Kosuge, Phys. Rev. Lett.  \textbf{79}, 3258 (1997).
		\bibitem{r8} H. Kageyama, K. Yoshimura, K. Kosuge, H. Nojiri, K. Owari and M. Motokawa, Phys. Rev. B \textbf{58}, 11150 (1998).
		\bibitem{r9} A. Jain, Sher Singh, and S. M. Yusuf, Phys. Rev. B \textbf{74}, 174419 (2006) and I. Nowik, A. Jain, S. M. Yusuf and J. V. Yakhmi, Phys. Rev. B \textbf{77}, 054403 (2008). and A. Jain, Sher Singh, and S. M. Yusuf, Phys. Rev. B \textbf{74}, 174419 (2006).
		\bibitem{r10} D. Flahaut, A. Maignan, S. H\'{e}bert, C. Martin, R. Retoux, and V. Hardy, Phys. Rev. B \textbf{70}, 094418 (2004).
		\bibitem{r11} Z. W. Ouyang, N. M. Xia, Y. Y. Wu, S. S. Sheng, J. Chen, Z. C. Xia L. Li and G. H. Rao, Phys. Rev. B \textbf{84}, 054435 (2011).
		\bibitem{r12} H. Kageyama, K. Yoshimura and K. Kosuge, J. Solid State Chem. \textbf{140}, 14 (1998).
		\bibitem{r13} S. Rayaprol, K. Sengupta and E. V. Sampathkumaran, Phys. Rev. B \textbf{67}, 180404(R) (2003) and E. V. Sampathkumaran and A. Niazi, Phys. Rev. B \textbf{65}, 180401(R) (2002).
		\bibitem{r14} G. Allodi, R. D. Renzi, S. Agrestini, C. Mazzoli and M. R. Lees, Phys. Rev. B \textbf{83}, 104408 (2011).
		\bibitem{r15} T. Burnus, Z. Hu, M. W. Haverkort, J. C. Cezar, D. Flahaut, V. Hardy, A. Maignan, N. B. Brookes, A. Tanaka, H. H. Hsieh, H.-J. Lin, C. T. Chen and L. H. Tjeng, Phys. Rev. B \textbf{74}, 245111 (2006).
		\bibitem{r16} H. Wu, M. W. Haverkort, Z. Hu, D. I. Khomskii, and L. H. Tjeng,  Phys. Rev. Lett.  \textbf{95}, 186401 (2005).
		\bibitem{r17} R. Fresard, C. Laschinger, T. Kopp and V. Eyert, Phys. Rev. B \textbf{69}, 140405(R) (2004).
		\bibitem{r18} S. Aasland, H. Fjellvlg and B. Haubackb, Solid State Commun. \textbf{101}, 187 (1997).
		\bibitem{r19} H. Kageyama, K. Yoshimura, K. Kosuge, H. Mitamura and T. Goto, J. Phys. Soc. Jpn. \textbf{66}, 1607 (1997) and H. Kageyama, K. Yoshimura, K. Kosuge, X. Xu and S. Kawano, J. Phys. Soc. Jpn. \textbf{67}, 357 (1998).
		
		\bibitem{r20} S. Agrestini, L. C. Chapon, A. Daoud, Aladine, J. Schefer, A. Gukasov, C. Mazzoli, M. R. Lees and O. A. Petrenko, Phys. Rev. Lett. \textbf{101}, 097207 (2008).
		\bibitem{r21} J. A. M. Paddison, S. Agrestini, M. R. Lees, C. L. Fleck, P. P. Deen, A. L. Goodwin, J. R. Stewart and O. A. Petrenko, Phys. Rev. B \textbf{90}, 014411 (2014).
		\bibitem{r22} G. Allodi, P. Santini, S. Carretta, S. Agrestini, C. Mazzoli, A. Bombardi, M. R. Lees and R. De Renzi, Phys. Rev. B  \textbf{89}, 104401 (2014).
		\bibitem{r23} L. C. Chapon, Phys. Rev. B  \textbf{80}, 172405 (2009).
		\bibitem{r24} Y. Kamiya and C. D. Batista,  Phys. Rev. Lett. \textbf{109}, 067204 (2012) and references therein.
		
		\bibitem{r25} S. De and A. Banerjee, J. Magn. Magn. Mater. \textbf{539}, 168349 (2021).
		\bibitem{r26} V. Hardy, M. R. Lees, O. A. Petrenko, D. M. Paul, D. Flahaut, S. H\'{e}bert and A. Maignan, Phys. Rev. B \textbf{70}, 064424 (2004) and V. Hardy, D. Flahaut, M. R. Lees and O. A. Petrenko, Phys. Rev. B \textbf{70}, 214439 (2004).
		\bibitem{r27} A. Maignan, V. Hardy, S. Hebert, M. Drillon, M. R. Lees, O. A. Petrenko, D. Mc K. Paulc and D. Khomskii, J. Mater. Chem.  \textbf{14}, 1231 (2004). 
		\bibitem{r28} C. H. Kim, K. H. Kim, S. H. Park, Hyun-jong Paik, J. H. Cho, and Bog G. Kim, J. Phys. Soc. Jpn. \textbf{74}, 2317 (2005).
		\bibitem{r29} J. Arai, H. Shinmen, S. Takeshita, T. Goko, J. Magn. Magn. Mater. \textbf{272}, 809 (2004).
		\bibitem{r30} A. Maignan, C. Michel, A. C. Masset, C. Martin, and B. Raveau, Eur. Phys. J. B  \textbf{15}, 657 (2000).
		\bibitem{r31} S. De, K. Kumar, A. Banerjee and P. Chaddah, AIP Conf. Proc. \textbf{1731}, 130026 (2016).
		\bibitem{r32} S. De and A. Banerjee, Physica B \textbf{572}, 125 (2019).
		\bibitem{r33} T. Kurata and H. Kawamura, J. Phys. Soc. Jpn. \textbf{64}, 232 (1995).
		
		\bibitem{r34} S. Agrestini, C. Mazzoli, A. Bombardi, and M. R. Lees, Phys. Rev. B \textbf{77}, 140403 (R) (2008).
		\bibitem{r35} S. Agrestini, C. L. Fleck, L. C. Chapon, C. Mazzoli, A. Bombardi, M. R. Lees and O. A. Petrenko, Phys. Rev. Lett. \textbf{106}, 197204 (2011).
		\bibitem{r36} C. L. Fleck, M. R. Lees, S. Agrestini, G. J. McIntyre and O. A. Petrenko, EPL  \textbf{90}, 67006 (2010).
		\bibitem{r37} A. Jain and S. M. Yusuf, Phys. Rev. B \textbf{83}, 184425 (2011).
		\bibitem{r38} P. Lampen, N. S. Bingham, M. H. Phan, H. Srikanth, H. T. Yi and S. W. Cheong, Phys. Rev. B \textbf{89}, 144414 (2014).
		
		\bibitem{r39} M. Tabuchi, S. Tsutsui, C. Masquelier, R. Kanno, K. Ado, I. Matsubara, S. Nasu and H. Kageyama, J. Solid State Chem. \textbf{140}, 159 (1998).
		\bibitem{r40} S. Niitaka, K. Yoshimura, K. Kosuge, K. Mibu, H. Mitamura and T. Goto, J. Magn. Magn. Mater. \textbf{260}, 48 (2003).
		
		\bibitem{r41} A. Bajpai and A. Banerjee, Rev. Sci. Instrum. \textbf{68}, 4075 (1997) and B. Dutta, K. Kumar, N. Ghodke and A. Banerjee, Rev. Sci. Instrum. \textbf{91}, 123905 (2020).
		
		\bibitem{r42} S. B. Roy, J. Phys.: Condens. Matter \textbf{25}, 183201 (2013).
		\bibitem{r43} A. Banerjee, K. Mukherjee, K. Kumar, and P. Chaddah, Phys. Rev. B \textbf{74}, 224445 (2006).
		\bibitem{r44} S. B. Roy, M. K. Chattopadhyay, P. Chaddah, J. D. Moore, G. K. Perkins, L. F. Cohen, K. A. Gschneidner, Jr. and V. K. Pecharsky, Phys. Rev. B \textbf{74}, 012403 (2006).
		\bibitem{r45} K. Kumar, A. K. Pramanik, A. Banerjee, S. B. Roy, S. Park, C. L. Zhang and S.-W. Cheong, Phys. Rev. B \textbf{73}, 184435 (2006).
		\bibitem{r46} A. Banerjee, K. Kumar and P. Chaddah, J. Phys.: Condens. Matter \textbf{21}, 026002 (2009).
		\bibitem{r47} A. Banerjee, A. K. Pramanik, K. Kumar and P. Chaddah, J. Phys.: Condens. Matter \textbf{18}, L605 (2006).
		\bibitem{r48} P. Chaddah, K. Kumar, and A. Banerjee, Phys. Rev. B \textbf{77}, 100402(R) (2008).
		
		
		
		\bibitem{r49} V. Hardy, C. Martin, G. Martinet and G. Andr\'{e}\, Phys. Rev. B \textbf{74}, 064413 (2006).
		
		\bibitem{r50} K. Binder and A. P. Young, Rev. Mod. Phys., \textbf{58}, 801 (1986).
		
		
		\bibitem{r51} A. Bajpai and A. Banerjee, Phys. Rev. B \textbf{62}, 8996 (2000).
		\bibitem{r52} A. K. Pramanik and A. Banerjee, Phys. Rev. B \textbf{81}, 024431 (2011).
		
	\end{thebibliography}
\end{document}